# INVESTIGATING OF LONGITUDINAL DEVELOPMENT PARAMETERS THROUGH AIR SHOWER SIMULATION BY DIFFERENT HADRONIC MODELS

AL-RUBAIEE A.AHMED & AHMED JUMAAH

Al-Mustansiriyah University, College of Science, Department of Physics,

Baghdad, Iraq

## Abstract

In this work the simulation of the Extensive Air Showers was performed by investigating the longitudinal development parameters ($N$ and $X_{max}$) by using a system for air shower simulation which is called AIRES version 2.6.0 at the energy range ($10^{14}$-$10^{19}$eV) for different primary particles like (gamma, electron, positron, proton and iron nuclei) and different zenith angles. The comparison of simulated longitudinal profile was fulfilled for different hadronic models (SIBYLL, QGSJET99 and SIBYLL S16).

## Introduction

When high-energy cosmic ray particles penetrate the Earth atmosphere they interact and generate a cascade of secondary particles, that is called Extensive Air Showers (EAS), but the only way to get information on these particles is to study EAS cascades which they initiate [1]. When a cosmic ray enters the atmosphere it collides with the nucleus of an air atom, producing a number of secondary particles. These go on to make further collisions, and the number of particle grows. In the end, the energy of the shower particles is cut down to the point where ionization losses to overcome and their number start to descend [2]. The longitudinal development of EAS is depended on the energy and type of the incident primary particle [3]. One of its characteristics, the atmospheric depth of shower maximum ($X_{max}$), which is often used to reconstruct the elemental composition of primary cosmic rays. The number of charged particles (N) in EAS as a function of atmospheric depth is intimately related to the type and energy of the primary particle which can be simulated by AIRES code [4]. A realistic air shower simulation system which includes electromagnetic interaction algorithms and links to the hadronic models SIBYLL [5] and QGSJET [6].

In this work the investigation of the longitudinal development profiles in EAS was performed by using AIRES code for different hadronic models like (SIBYLL [5], QGSJET99 [6] and SIBYLL S1.6 [7]) packages for the simulation of hadronic processes. The simulation was performed for different primary particles (gamma, electron, positron, proton and iron nuclei) at the high energy range ($10^{14}$-$10^{19}$ eV) for different zenith angles (0°, 10° and 30°).

## Longitudinal Profile

The Gaisser-Hillas formula gives the approximate number of charged particles as a function of atmospheric depth along the shower axis [8]:

$$N(X) = N_{max} \left(\frac{X}{X_{max}}\right)^{X_{max}/\lambda} \exp\left(X_{max} - X\right)/\lambda, \qquad (1)$$

where $X$ is the atmospheric depth (in g cm$^{-2}$); $N_{max}$ is the number of charged particles; $X_{max}$ is the depth of shower maximum and $\lambda$ is a characteristic length parameter (a scale constant with a value 70 g cm$^{-2}$) [9]. This is a Gamma function form which naturally arises in cascade theory, and assumes that the first interaction is at $X=0$. The first interaction point is the location of initial collision of the cosmic ray particle with atmosphere. A forth parameter is often introduced into Eq. (1), ostensibly to allow for a variable first interaction point [10]:

$$N(X) = N_{max} \left(\frac{X - X_0}{X_{max} - X_0}\right)^{(X_{max}-X_0)/\lambda} \exp\left(X_{max} - X\right)/\lambda. \qquad (2)$$

Where $X_0$ is the depth of the point of first interaction. The value of $X_0$ depends on the collision cross section and hence the energy and mass composition of the particle. The depth of shower maximum $X_{max}$ depends on the position of $X_0$, the shower energy and composition. But one can express the shower longitudinal development as function of the shower age *s* that defined as $s=3X/X+2X_{max}$ instead of depth $X$, by using *s*. Translating the depth $X_{max}$ into age *s* and using the normalized shower size $n=N/N_{max}$, the Eq. (2) becomes [11]:

$$n(s) = \left(1 - \frac{(1-s)3T_m}{(3-s)(T_m - T_o)}\right)^{T_m - T_o} e^{3T_m(1-s/3-s)} \quad (3)$$

where $T_m = X_{max}/\lambda$ and $T_0 = X_0/\lambda$.

## Results and Discussions`

In this work we apply the AIRES simulation code to study generation of primary particles like (gamma, electron, positron, proton and iron nuclei) induced air showers at primary energy ($10^{14}$ to $10^{19}$eV) and explores the longitudinal development. Although the simulation of air showers at the lowest energy is not a very realistic application we intend in the present work to investigate the longitudinal development by the different hadronic models (SIBYLL, QGSJET99 and SIBYLL S16) at the highest energy ranges ($10^{14}$-$10^{19}$eV) for different zenith angles. All simulations performed by the AIRES code were obtained by using the thinning factor $10^{-4}$ relative thinning case.

Figure (1) shows the longitudinal development for the primary particles (gamma, electron and proton), when air shower curves represent number of particles with fixed atmospheric depth (in g cm$^{-2}$). The results of simulation are compared between different hadronic models at the energy range ($10^{14}$-$10^{19}$eV) for vertical showers.

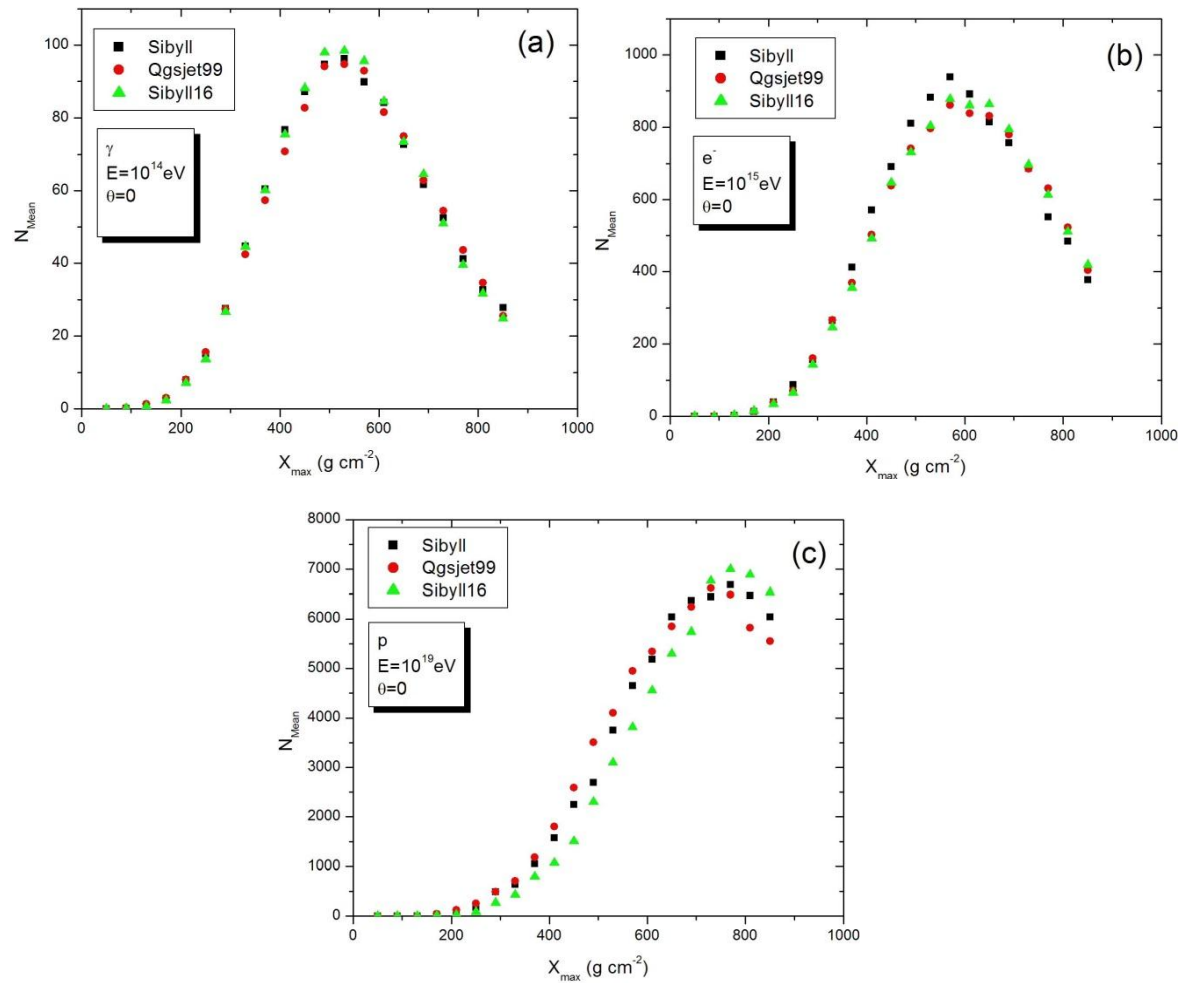

Figure (1): The longitudinal development simulated by AIRES code for vertical shower and different hadronic model (SIBYLL, QGSJET99 and SIBYLL S16), for different primary particles: (a) gamma particle at the energy ($10^{14}$eV); (b) primary electron at the energy ($10^{15}$eV); (c) primary proton at the energy ($10^{19}$eV).

The longitudinal development for the primary particles (electron, positron and iron nuclei) which displayed in figure (2) reflects another importance or characteristic of the EAS development that simulated by AIRES code for $\theta =10°$, and the fluctuations are very large for $10^{-4}$ relative thinning level, that they reduce immediately when the thinning is lowered.

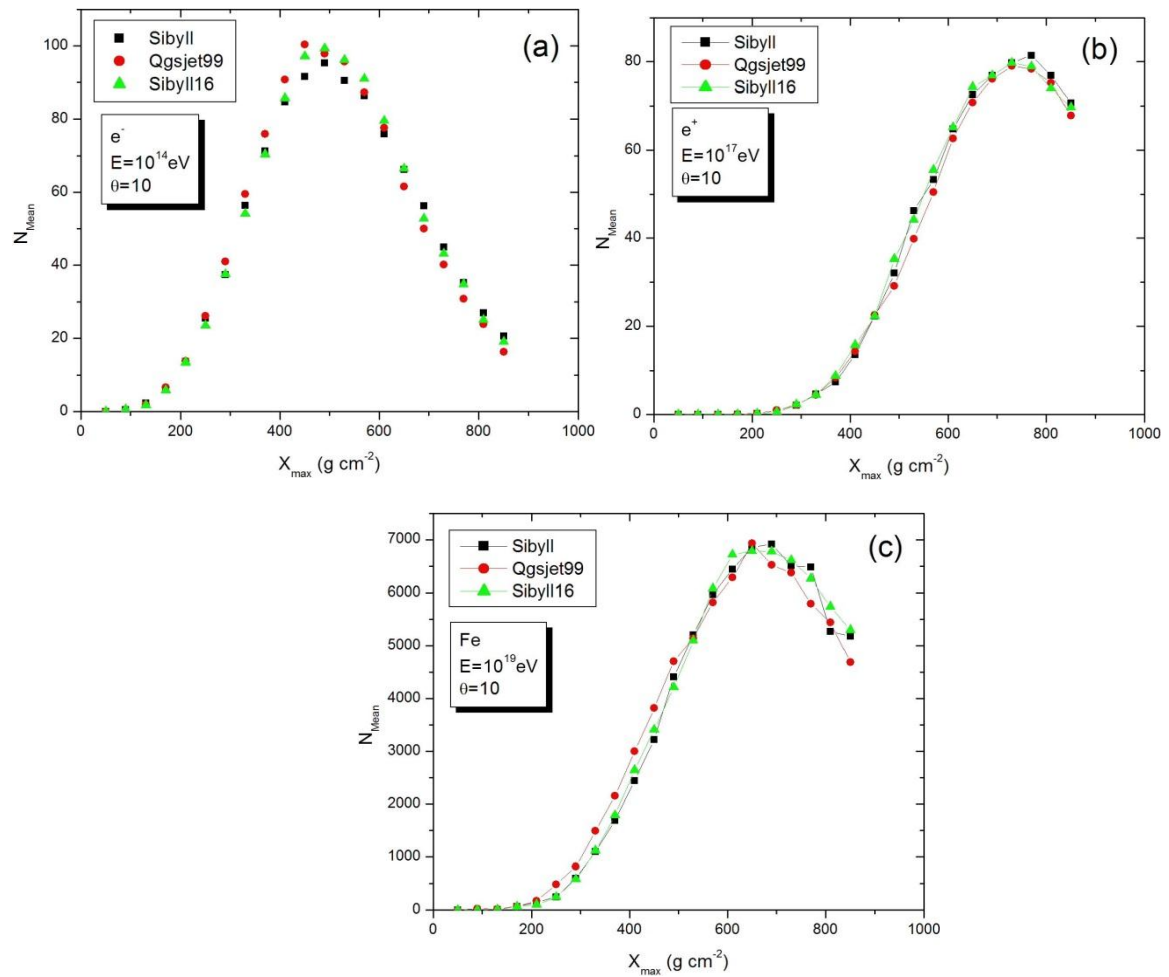

Figure (2): The longitudinal development simulated by AIRES code for $\theta=10°$ and different hadronic models (SIBYLL, QGSJET99 and SIBYLL S16), for different primary particles: (a) primary electron at the energy ($10^{14}$eV); (b) primary positron at the energy ($10^{17}$eV); (c) iron nuclei at the energy ($10^{19}$eV).

The figure (3) Shows the results of simulated longitudinal development in EAS for different primary particles (proton, gamma, iron nuclei), at the energy range ($10^{14}$-$10^{19}$eV), for $\theta$ =30°. Through this figure one can see the effect of the hadronic models on the fluctuations of the longitudinal development of primary particles, in the same conditions as in figure (2) for different hadronic models.

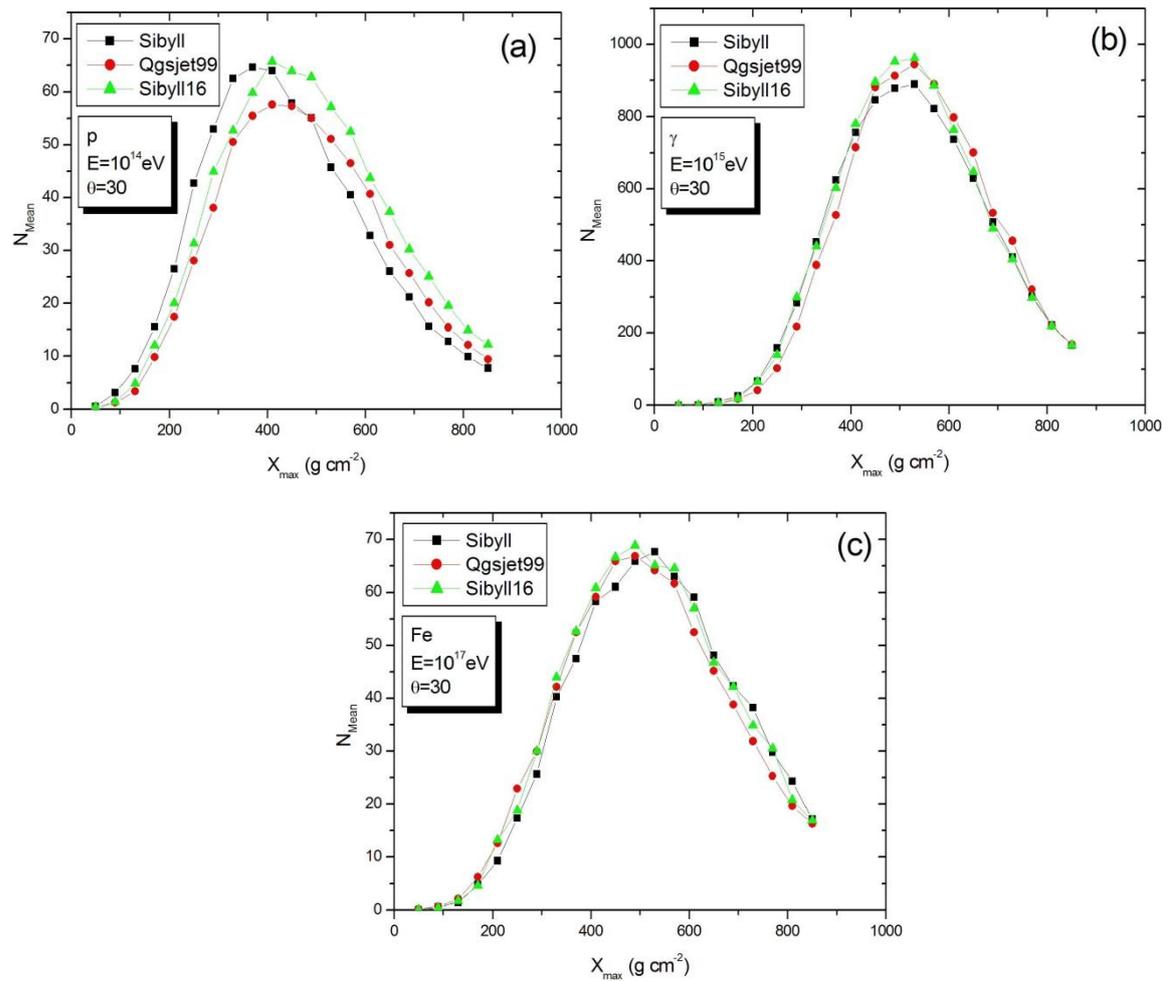

Figure (3): The longitudinal development simulated by AIRES code for $\theta$=30° and different hadronic models (SIBYLL, QGSJET99 and SIBYLL S16), for different primary particles: (a) primary proton at the energy ($10^{14}$eV); (b) gamma particle at the energy ($10^{15}$eV); (c) iron nuclei at the energy ($10^{17}$eV).

## Conclusion

In the present work the code AIRES was used for simulating the development of air showers in the atmosphere initiated by different primary particles (gamma, electron, positron, proton and iron nuclei) at the energy range ($10^{14}$-$10^{19}$eV) with different zenith angles. The simulation of the longitudinal development has shown an opportunity of primary particle identification and knowledge of the atmospheric depth for the cosmic ray spectrum. The obtained results led to the conclusions that: (a) Found difference between calculations in different hadronic models; (b) Showers induced by heavy particles are more sensitive to atmospheric depth maximum. The main advantage of the given approach consists possibility of a longitudinal development analysis of real events which detected with extensive air shower array and reconstruction of the primary cosmic ray energy spectrum and mass composition.